\documentstyle[12pt,epsfig]{article}
\begin{document}

\vspace{0.2cm}

\begin{center}
{\large\bf Neutrino Decays and Neutrino Electron Elastic Scattering
\\ in Unparticle Physics}
\end{center}

\vspace{.5cm}
\begin{center}
{\bf Shun Zhou} \footnote{E-mail: zhoush@mail.ihep.ac.cn} \\
{\small\sl Institute of High Energy Physics, Chinese Academy of
Sciences, Beijing 100049, China}
\end{center}

\vspace{3.5cm}

\begin{abstract}
Following Georgi's unparticle scheme, we examine the effective
couplings between neutrinos and unparticle operators. As an
immediate consequence, neutrinos become unstable and can decay into
the unparticle stuff. Assuming the dimension transmutation scale is
around $\Lambda^{}_{\cal U} \sim 1 ~{\rm TeV}$, we implement the
cosmological limit on the neutrino lifetime to constrain the
neutrino-unparticle couplings for different scaling dimensions $d$.
In addition, provided that the electron-unparticle coupling is
restricted due to the precise measurement of the anomalous magnetic
moment of electron, we calculate the unparticle contribution to the
neutrino-electron elastic scattering. It is more important to
jointly deal with the couplings of the unparticle to the standard
model particles rather than separately. Taking into account both
electron- and neutrino-unparticle couplings, we find that the
scaling dimension of the scalar unparticle should lie in the narrow
range $1 < d < 2$ by requiring the observables to be physically
meaningful. However, there is no consistent range of $d$ for the
vector unparticle operator.
\end{abstract}

\newpage

\section{Introduction}

It has been shown by Banks and Zaks \cite{BZ} that the non-Abelian
gauge theories with massless fermions can have an infrared-stable
fixed point. However, this kind of scale-invariant theory requires
the non-integral number of fermion generations and thus is not
realized by nature. Recently, Georgi has pointed out that Banks and
Zaks (BZ) fields and the standard model (SM) fields may coexist at
some high energy scale, where the interaction between these two sets
of fields is mediated by the messenger field with the mass scale
$M^{}_{\cal U}$ \cite{Georgi}. At the energy scale lower than
$M^{}_{\cal U}$, physical phenomena are described by the
non-renormalizable operators, which are suppressed by the inverse
powers of $M^{}_{\cal U}$ and of the form $\lambda {\cal O}^{}_{\rm
SM} {\cal O}^{}_{\rm BZ}/M^k_{\cal U}$ just as in the conventional
effective theories. Note that $\lambda$ is the dimensionless
coupling constant, ${\cal O}^{}_{\rm BZ}$ and ${\cal O}^{}_{\rm SM}$
are respectively the operators composed of BZ and SM fields. It is
well known that the radiative corrections in the scale-invariant
theory will induce the dimension transmutation \cite{Coleman}, which
means that an energy scale $\Lambda^{}_{\cal U}$ appears even if
there is only one dimensionless coupling in the generic theory of BZ
fields. As argued by Georgi \cite{Georgi}, below the scale
$\Lambda^{}_{\cal U}$, the BZ field operator ${\cal O}^{}_{\rm BZ}$
should match onto the unparticle operator ${\cal O}^{}_{\cal U}$
with a non-integral scaling dimension $d$. Therefore, we have the
low-energy operators $\lambda^\prime \Lambda^{(d^{}_{\rm BZ}
-d)}_{\cal U} {\cal O}^{}_{\rm SM} {\cal O}^{}_{\cal U}/M^k_{\cal
U}$, where $d^{}_{\rm BZ}$ is the scaling dimension of ${\cal
O}^{}_{\rm BZ}$.  In such a setup, Georgi has further claimed that
the unparticle effects can show up at the colliders as the missing
energy and may be promising to be discovered prior to the other new
physics beyond the SM \cite{Georgi}.

Shortly after Georgi's proposal, enormous studies have been
performed to investigate the unparticle phenomenology
\cite{Georgi2}-\cite{Mathews}. Since the interaction between the
unparticle and SM particles is unclear, one may introduce the
operator which can influence the processes well measured in
experiments. In this direction of thought, the invisible unparticle
$\cal U$ as the final state has been considered in the top quark
decay $t \to u + {\cal U}$ \cite{Georgi}, the electron-positron
annihilation $e^+ + e^- \to \gamma + {\cal U}$ and the hadronic
processes, such as $q + q \to g + {\cal U}$ \cite{Georgi2,Cheung}.
The importance of the interference between the SM and
unparticle-induced contributions to a specific process is
highlighted in Ref. \cite{Georgi2}, where the typical channel $e^+ +
e^- \to \mu^+ + \mu^-$ is analyzed in detail. In some sense, the
unparticle sector serves as one kind of new physics beyond the SM.
One should take into account the unparticle effects on all the
familiar processes. Bearing this in mind, some authors have
discussed the possible new origin of CP violation \cite{Geng}, the
deep inelastic scattering \cite{Ding}, the anomalous magnetic moment
of charged leptons $(g-2)$ \cite{Cheung,Luo,Liao} and
lepton-flavor-violating processes \cite{Aliev,Wang} in unparticle
physics. On the other hand, since the Lorentz group representations
that the unparticle operators belong to are not restricted, they can
be of scalar, vector \cite{Georgi2,Cheung} or spinor types
\cite{Luo}. It is also natural to assume that the unparticle
operator is invariant under the gauge symmetry of the SM, then one
can systematically write down the gauge invariant effective
operators as in Ref. \cite{He}.

However, the couplings between neutrinos and the unparticle have not
yet been touched thus far. Thanks to the elegant neutrino
oscillation experiments, we are now convinced that neutrinos are
massive and lepton flavors are mixed \cite{Review}. Massive
neutrinos play an important role in astrophysics and cosmology, for
instance, the energy density of active neutrinos may affect the
light element abundance in the big bang nucleosynthesis scenario and
the cosmic microwave backgroud. If the unparticle sector couples to
neutrinos, heavier neutrinos can decay into the light ones and the
invisible unparticle stuff as we will show later. At present, the
most stringent limit on the neutrino lifetime comes from the solar
neutrino experiment \cite{Beacom}: $\tau/m \geq 10^{-4}~ {\rm s ~
eV^{-1}}$ with $m$ and $\tau$ being the mass and lifetime of
neutrinos. The detection of the decay of neutrinos from other
astrophysical sources, such as a Galactic supernova or a distant
Active Galactic Nuclei, may improve the constraint by several orders
of magnitude. It has been recently proposed \cite{Serpico} that
future cosmological observations can measure the sum of neutrino
masses to the accuracy about $10^{-2}~{\rm eV}$, thus they may serve
as the best probe of the neutrino lifetime $\tau/m \geq 10^{16} ~
m^{-5/2}_{50} ~ {\rm s ~ eV^{-1}}$, where $m^{}_{50} \equiv
m/(50~{\rm meV})$. Obviously, this bound is more serious than that
from solar neutrino analysis and can be used to constrain the
neutrino-unparticle couplings. Note that we will concentrate on the
non-radiative decays of neutrinos, the cosmological limit on the
radiative decays can be found in \cite{Serpico1}.

This paper is organized as follows. In Sec. 2, we introduce the
interaction between neutrinos and the scalar unparticle operator in
addition to the electron-unparticle coupling. The latter is
restricted by the precise measurement of the anomalous magnetic
moment of electron, while the former is constrained from the bound
on neutrino lifetime. Furthermore, we also calculate the cross
section of neutrino-electron elastic scattering, in which these two
kinds of couplings simultaneously appear. It is found that the
scaling dimension of the scalar unparticle operator, which couples
with both electrons and neutrinos, should stay in the range $1 < d <
2$, since the observables should be physically meaningful. In
comparison, the vector unparticle operator is considered in Sec. 3,
where we show that there is no consistent range of the scaling
dimension. Therefore, it is more important to jointly deal with the
unparticle effects in different physical processes, in which the
common unparticle operator is present. Sec. 4 is devoted to the
summary of conclusions.

\section{Scalar Unparticle Operator}

The simplest case is to consider the scalar unparticle operator, and
our working effective Lagrangian takes the following form
\begin{eqnarray}
{\cal L}_S = \frac{\lambda^{\alpha \beta}_l}{\Lambda^{d-1}_{\cal
U}}\bar{l}^{}_\alpha l^{}_\beta {\cal O}^{}_{\cal U} +
\frac{\lambda^{\alpha \beta}_\nu}{\Lambda^{d-1}_{\cal
U}}\bar{\nu}^{}_\alpha \nu^{}_\beta {\cal O}^{}_{\cal U} \; + {\rm
h.c.},
\end{eqnarray}
where $\alpha, \beta = e, \mu, \tau$ are the flavor indices,
$\Lambda^{}_{\cal U}$ the dimension transmutation scale, $d$ the
scaling dimension of the scalar unparticle operator,
$\lambda^{}_l$'s and $\lambda^{}_\nu$'s the relevant coupling
constants. The most important consequence of the first term in Eq.
(1) is that the unparticle contributes to the anomalous magnetic
moments of charged leptons. On the other hand, because of its
flavor-violating feature, the scalar unparticle can mediate the
lepton-flavor-changing rare decays, such as $\mu^- \to e^- e^+ e^-$
\cite{Aliev}. In the following we will concentrate on the flavors in
the neutrino sector and thus only consider the electron-unparticle
coupling $\lambda^{ee}_l \equiv \lambda^{}_e$. Note that the
unparticle contribution to the anomalous magnetic moment of electron
has already been discussed in Refs. \cite{Georgi2,Cheung,Liao}:
\begin{eqnarray}
\Delta a^{}_e = -\frac{3A^{}_d \lambda^2_e}{16\pi^2 \sin(d\pi)}
\frac{\Gamma(2-d) \Gamma(2d-1)}{\Gamma(d+2)}
\left(\frac{m^2_e}{\Lambda^2_{\cal U}}\right)^{d-1} \; ,
\end{eqnarray}
where $m^{}_e = 0.51~{\rm MeV}$ is the electron mass and $A^{}_d$ is
a normalization constant defined in Eq. (4) below. The scaling
dimension should be $d < 2$ in order that the integral is finite. As
argued by Georgi \cite{Georgi2}, one can choose the theoretically
consistent values of $d$ in the range $1< d <2$. The difference
between the current experimental data and the SM prediction of
$a^{}_e$ is $|\Delta a^{}_e| \leq 15 \times 10 ^{-12}$ \cite{Liao},
which can place a strict constraint on the parameters
$\Lambda^{}_{\cal U}$, $d$ and $\lambda^{}_e$. As shown in Fig. 1,
the severest bound on the electron-unparticle coupling is
$\lambda^{}_e < 10^{-4}$ if $d = 1.1$, where the dimension
transmutation scale is typically taken to be $\Lambda^{}_{\cal U} =
1~{\rm TeV}$. Note that if $d$ gets larger values, the constraint on
$\lambda^{}_e$ can be relaxed.

We introduce the lepton-flavor-violating couplings of neutrinos to
the unparticle, which is well motivated by neutrino flavor mixing as
observed in the neutrino oscillations. From the second term in Eq.
(1), one can observe that heavier neutrinos become unstable and can
decay into the unparticle stuff and the light ones. Nevertheless,
the mass ordering of neutrinos is not uniquely determined. For
simplicity, we assume that the lightest one is massless: for the
normal mass hierarchy, $m^{}_1 = 0$, $m^{}_2 \approx 9.0 ~{\rm meV}$
and $m^{}_3 \approx 50~{\rm meV}$; for the inverted mass hierarchy,
$m^{}_3 =0 $ and $m^{}_1 \approx m^{}_2 \approx 50~{\rm meV}$. It is
more convenient to work in the neutrino mass eigenstate basis, which
is defined as $\nu^{}_i = \sum_\alpha V^*_{\alpha i} \nu^{}_\alpha$
with $V$ being the neutrino mixing matrix. In this basis, the
interaction between neutrinos and the unparticle can be written as
$\lambda^{ij}_\nu\bar{\nu}^{}_i \nu^{}_j {\cal O}^{}_{\cal
U}/\Lambda^{d-1}_{\cal U}$ with $\lambda^{ij}_\nu \equiv
\sum_{\alpha, \beta} V^*_{\alpha i} \lambda^{\alpha \beta}_\nu
V^{}_{\beta j}$. According to the scale invariance in the unparticle
sector, we have \cite{Georgi2}
\begin{eqnarray}
\int {\rm d}^4 x  \langle0|T[{\cal O}^{}_{\cal U}(x) {\cal
O}^\dagger_{\cal U}(0)]|0\rangle e^{ipx} = i \frac{A^{}_d}{2}
\frac{1}{\sin(d\pi)} (-p^2 -i\epsilon)^{d-2} \; ,
\end{eqnarray}
while there is an additional Lorentz factor $(-g^{\mu \nu} + p^\mu
p^\nu/p^2)$ for the vector unparticle operator ${\cal O}^\mu_{\cal
U}$, which satisfies the transverse condition $\partial^{}_\mu {\cal
O}^\mu_{\cal U} = 0$. The normalization constant is defined as
\begin{eqnarray}
A^{}_d \equiv \frac{16 \pi^{5/2}}{(2\pi)^{2d}}
\frac{\Gamma(d+1/2)}{\Gamma(d-1) \Gamma(2d)} \; .
\end{eqnarray}
It is straightforward to figure out the differential decay rate of
neutrinos, namely the process $\nu^{}_j (p,s^{}_1) \to \nu^{}_i (k,
s^{}_2) + {\cal U}(q)$,
\begin{eqnarray}
{\rm d}\Gamma^{}_j = \frac{1}{2m^{}_j} (2\pi)^4 \delta^4(p-k-q)
\left|{\cal M} \right|^2 \frac{{\rm d}^3 k}{(2\pi)^3
2k^0}\left[A^{}_d \theta(q^0) \theta(q^2) (q^2)^{d-2} \frac{{\rm
d}^4 q}{(2\pi)^4}\right] \; ,
\end{eqnarray}
where the invariant matrix element $\left|{\cal M} \right|^2 =
2\left|\lambda^{ij}_\nu\right|^2 k\cdot p/\Lambda^{2(d-1)}_{\cal U}$
has been summed over the final state spins and averaged over the
initial state spin. After integrating over the phase space, we get
\begin{eqnarray}
\frac{{\rm d}\Gamma^{}_j}{{\rm d}E^{}_i} = \frac{A^{}_d
\left|\lambda^{ij}_\nu\right|^2}{4\pi^2 \Lambda^{2(d-1)}_{\cal U}}
\frac{E^2_i\theta(m^{}_j - 2E^{}_i)}{(m^2_j -2m^{}_j E^{}_i)^{2-d}}
\; ,
\end{eqnarray}
where $\nu^{}_i$ is the lightest neutrino and its mass has been set
to be vanishing for simplicity, and $E^{}_i$ is the energy of the
final state neutrino. The total decay rate is given by
\begin{eqnarray}
\Gamma^{}_j = \int^{m^{}_j/2}_0 \left(\frac{{\rm d}\Gamma^{}_j}{{\rm
d}E^{}_i}\right) {\rm d}E^{}_i = \frac{A^{}_d
\left|\lambda^{ij}_\nu\right|^2}{16 \pi^2 d(d^2-1)}
\left(\frac{m^2_j}{\Lambda^2_{\cal U}}\right)^{d-1} m^{}_j \; .
\end{eqnarray}
Combining the expression of $a^{}_e$ and the above equation, we see
that the scaling dimension should lie in the range $1 < d < 2$ in
order that these physical quantities are well defined. To make clear
the dependence of the differential decay rate on $d$, we can define
\begin{eqnarray}
m^{}_j \frac{{\rm d} \ln \Gamma^{}_j}{{\rm d}E^{}_i} = 4
d(d^2-1)(1-2y)^{d-2}y^2 \;,
\end{eqnarray}
where $y \equiv E^{}_i/m^{}_j <  1/2$. This is the same as the
process $t \to u + {\cal U}$ considered in \cite{Georgi}, however,
the scaling dimension is now $1<d<2$ as restricted by the anomalous
magnetic moment of electron and the neutrino decay rate. It is
evident that the behavior of the decay rate with the unparticle
stuff in the final state is drastically different from the ordinary
two-body decay case. Since it is almost impossible to measure decay
products of neutrinos unlike the decay of top quark \cite{Georgi},
the most important and relevant quantity is the total decay rate
$\Gamma^{}_j$, or equivalently the neutrino lifetime $\tau^{}_{\cal
U} \equiv \Gamma^{-1}_j$. From Eq. (7), we can obtain
\begin{eqnarray}
\frac{\tau^{}_{\cal U}}{m^{}_j}= \frac{16 \pi^2 d(d^2-1)}{A^{}_d
\left|\lambda^{ij}_\nu\right|^2} \left(\frac{\Lambda^2_{\cal
U}}{m^2_j}\right)^{d-1} \frac{1}{m^2_j} \; ,
\end{eqnarray}
which should be contrasted with the future cosmological constraint
$\tau/m \geq 10^{16} ~ m^{-5/2}_{50} ~ {\rm s ~ eV^{-1}} \approx
1.5\times 10^{31}~ m^{-5/2}_{50}~{\rm eV^{-2}}$. As is mentioned
before, the mass hierarchy of neutrinos is still undetermined.
However, the most crucial limit on $\lambda^{ij}_\nu$ is the case
with $m^{}_j = 50~{\rm meV}$ and $m^{}_i = 0$, for which the
numerical analysis is shown in Fig. 2. One can observe that the
neutrino-unparticle couplings are restricted to be on the order of
$10^{-5}$ for $d=1.1$ and $\Lambda^{}_{\cal U} = 1~{\rm TeV}$. This
bound will be relaxed when $d$ becomes larger, for instance,
$\lambda^{}_\nu \sim 0.5$ if $d=1.7$. Note that the constraint on
$\lambda^{ij}_\nu$ can be directly converted into that on
$\lambda^{\alpha \beta}_\nu$ by using the neutrino mixing matrix,
which is now measured in neutrino oscillation experiments to an
acceptable degree of accuracy. Roughly we expect them to be of the
same order.

Now we proceed to discuss the physical processes, in which both
electron- and neutrino-unparticle couplings are present. It is easy
to note that the unparticle will contribute to the neutrino-electron
elastic scattering. For the $\nu^{}_e e^-$ elastic scattering, the
unparticle contribution will interfere with the charged- and
neutral-current amplitudes, while for the $\nu^{}_\alpha e^-$
($\alpha = \mu, \tau$) interference between the unparticle and
neutral-current components arises. The relevant neutrino-unparticle
couplings $\lambda^{\alpha \alpha}_\nu$ in these two cases can be
probed by measuring the cross sections of neutrino electron elastic
scattering. However, the $\nu^{}_\alpha e^- \to \nu^{}_\beta e^-$
for $\alpha \neq \beta$ can not occur in the SM, and the couplings
are also relevant to the neutrino decays. So we will calculate this
flavor-changing process, and point out its implication for the
unparticle physics. Note that this case is similar to the
non-standard interaction discussed in Refs. \cite{Lindner,Valle}.
The invariant matrix element can be computed for $\nu^{}_\alpha (k)
+ e^-(p) \to \nu^{}_\beta (k^\prime) + e^-(p^\prime)$ scattering:
\begin{eqnarray}
\frac{1}{4} \sum_{s} |{\cal M}|^2 = \frac{1}{16} \left[\frac{A^{}_d
\lambda^{}_e \lambda^{\alpha \beta}_\nu}{\Lambda^{2(d-1)}_{\cal U}
\sin(d\pi)} \right]^2 (k \cdot k^\prime) (p \cdot p^\prime)
\left[-(k-k^\prime)^2 - i\epsilon \right]^{2(d-2)}\; ,
\end{eqnarray}
where we have summed over the final spins and averaged over the
initial spins. The total cross section in the center-of-mass
reference frame is given by
\begin{eqnarray}
\sigma(s) = \int \frac{1}{4(k \cdot p)} (2\pi)^4 \delta^4(k+p -
k^\prime -p^\prime) \left(\frac{1}{4} \sum_{s} |{\cal M}|^2 \right)
\frac{{\rm d}^3 k^\prime}{(2\pi)^3 2k^\prime_0} \frac{{\rm d}^3
p^\prime}{(2\pi)^3 2p^\prime_0} \; ,
\end{eqnarray}
where the lepton masses have been neglected. A straightforward
calculation leads to the differential cross section
\begin{eqnarray}
\frac{{\rm d}\sigma(s)}{{\rm d}\cos \theta} = \frac{1}{32\pi \cdot
4^d} \left[\frac{A^{}_d \lambda^{}_e \lambda^{\alpha
\beta}_\nu}{\Lambda^{2(d-1)}_{\cal U} \sin(d\pi)} \right]^2 s^{2d-3}
(1-\cos \theta)^{2(d-1)} \; ,
\end{eqnarray}
where $s=(k+p)^2$ is the center-of-mass energy square and $\theta$
is the azimuthal angle. Note that the total cross section $\sigma(s)
\propto (s/\Lambda^2_{\cal U})^{2(d-1)}/[64\pi(2d - 1)s]$ is always
regular for $1 < d < 2$. Given the information about $\lambda^{}_e$
and $\lambda^{\alpha \beta}_\nu$, one can predict the total cross
section of the flavor-changing neutrino-electron scattering. For
example, we take the scaling dimension $d = 1.7$ and
$\Lambda^{}_{\cal U} = 1 ~ {\rm TeV}$, then $\lambda^{}_e \leq 1.0$
and $\lambda^{\alpha \beta}_\nu \leq 0.5$ as respectively indicated
by Fig. 1 and Fig. 2. Finally we get $\sigma \leq 1.4 \times
10^{-36} ~{\rm cm^2}$ for $\sqrt{s} = 200~{\rm GeV}$, which should
be compared with the SM prediction of the neutrino-electron elastic
scattering. Note that the present limit on neutrino lifetimes is
just $\tau/m \geq 10^{-4}~{\rm s~eV^{-1}}$, which will hardly
constrain $\lambda^{}_\nu$. In this case, we may inversely use the
experimental data on extra contributions to neutrino-electron
elastic scattering to extract the information about
neutrino-unparticle couplings.

\section{Vector Unparticle Operator}
If the vector unparticle operator couples both to charged-leptons
and to neutrinos, the Lagrangian can be written as
\begin{eqnarray}
{\cal L}_V = \frac{\lambda^{\alpha \beta}_l}{\Lambda^{d-1}_{\cal
U}}\bar{l}^{}_\alpha \gamma^{}_\mu l^{}_\beta {\cal O}^{\mu}_{\cal
U} + \frac{\lambda^{\alpha \beta}_\nu}{\Lambda^{d-1}_{\cal
U}}\bar{\nu}^{}_\alpha \gamma^{}_\mu \nu^{}_\beta {\cal
O}^{\mu}_{\cal U} \; + {\rm h.c.},
\end{eqnarray}
which will cause some interesting implications for unparticle
physics. For simplicity, we still focus on the neutrino flavors and
are just concerned about the electron-unparticle coupling. The
vector unparticle contribution to the anomalous magnetic moment of
electron has also been calculated in \cite{Cheung,Liao},
\begin{eqnarray}
\Delta a^{}_e = -\frac{A^{}_d \lambda^2_e}{8\pi^2 \sin(d\pi)}
\frac{\Gamma(3-d) \Gamma(2d-1)}{\Gamma(d+2)}
\left(\frac{m^2_e}{\Lambda^2_{\cal U}}\right)^{d-1} \; ,
\end{eqnarray}
where the requirement $d < 2$ should be satisfied. On the other
hand, one can analogously discuss the total rate of neutrino decays
into the vector unparticle stuff, i.e. $\nu^{}_j(p, s^{}_1) \to
\nu^{}_i(k, s^{}_2) + {\cal U}(q, {\cal S})$. The invariant matrix
element then is
\begin{eqnarray}
\frac{1}{2} \sum_{s^{}_1, s^{}_2, {\cal S}}\left|{\cal M}\right|^2 =
\frac{2\left|\lambda^{ij}_\nu\right|^2}{\Lambda^{2(d-1)}_{\cal
U}}\left[2(k\cdot q)(p\cdot q)/q^2 + k\cdot p\right] \; .
\end{eqnarray}
After substituting the above equation into Eq. (5) and integrating
over the phase space, we get
\begin{eqnarray}
\frac{{\rm d}\Gamma^{}_j}{{\rm d}E^{}_i} = \frac{A^{}_d
\left|\lambda^{ij}_\nu\right|^2}{4\pi^2 \Lambda^{2(d-1)}_{\cal U}}
\frac{m^{}_j E^2_i (3m^{}_j - 4 E^{}_i)}{(m^2_j -2m^{}_j
E^{}_i)^{3-d}} \theta(m^{}_j - 2E^{}_i) \; ,
\end{eqnarray}
where $E^{}_i$ is the energy of the final state neutrino. The total
decay rate is given by
\begin{eqnarray}
\Gamma^{}_j = \int^{m^{}_j/2}_0 \left(\frac{{\rm d}\Gamma^{}_j}{{\rm
d}E^{}_i}\right) {\rm d}E^{}_i = \frac{3 A^{}_d
\left|\lambda^{ij}_\nu\right|^2}{16 \pi^2 d(d-2)(d+1)}
\left(\frac{m^2_j}{\Lambda^2_{\cal U}}\right)^{d-1} m^{}_j \; .
\end{eqnarray}
In the above equation, $d > 2$ is demanded in order that the total
decay rate is finite and positive. Unfortunately, this condition on
the scaling dimension conflicts with that from the calculation of
$a^{}_e$. It seems very strange that the vector unparticle
interacting with electrons cannot simultaneously interact with
neutrinos in the way depicted in Eq. (13).

\section{Conclusions}

In summary, we have introduced effective couplings between neutrinos
and the scalar unparticle operator in addition to the
electron-unparticle coupling. Because of the neutrino-unparticle
interaction, heavier neutrinos become unstable and can decay into
the unparticle stuff. The decay rate of neutrinos has been
calculated and confronted with the cosmological limit on the
neutrino lifetime. Provided that the electron-unparticle coupling is
constrained from the precise measurement of the anomalous magnetic
moment of electron, and the neutrino-unparticle coupling from the
bound on the neutrino lifetime, we also figure out the cross section
of the lepton-flavor-changing neutrino-electron scattering. The
scaling dimension turns out to be in the range $1 < d < 2$ in order
that the anomalous magnetic moment of electron in Eq. (2) and the
neutrino decay rate in Eq. (7) are well defined. In comparison, we
also consider the effective interactions of electrons and neutrinos
with the vector unparticle operator. It is found that there is no
consistent region for the scaling dimension in this case. Therefore,
we remark that it is necessary to systematically consider the
physical processes in which the common unparticle operator couples
to the SM particles. This has been proved useful for the
determination of the scaling dimension.

\vspace{0.5cm}

The author would like to thank Professor Z.Z. Xing for stimulating
discussions, constant encouragement and reading the manuscript. He
is also grateful to W. Wang, W. Chao and H. Zhang for helpful
discussions, and particularly to J.F. Beacom, Z.T. Wei and G.H. Zhu
for useful comments. This work was supported in part by the National
Natural Science Foundation of China.

\newpage

\begin{figure}
\vspace{-12.0cm}
\epsfig{file=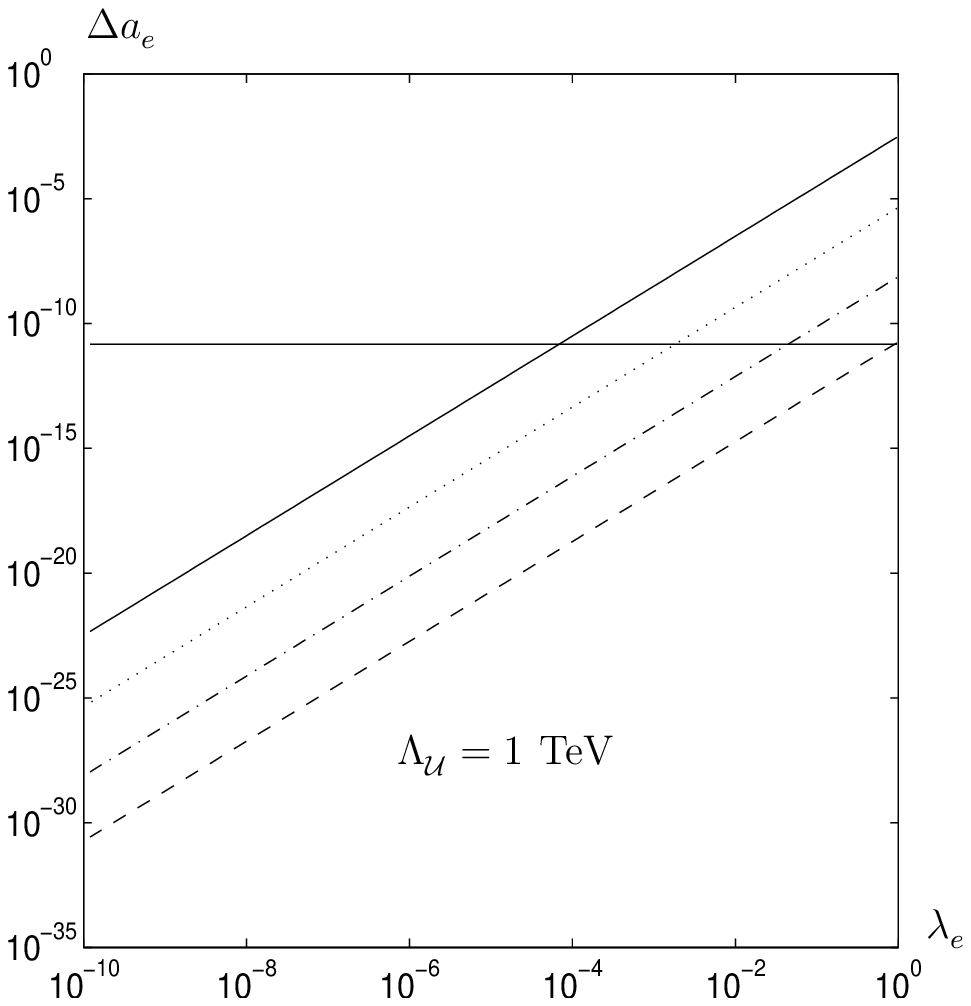,bbllx=0.5cm,bblly=14cm,bburx=12.5cm,bbury=34cm,%
width=10cm,height=16cm,angle=0,clip=0} \vspace{6.0cm}
\caption{Numerical illustration of $(\Delta a^{}_e, \lambda^{}_e)$
for different scaling dimensions of the scalar unparticle operator:
$d = 1.1$ (solid line), $d=1.3$ (dotted line), $d=1.5$
(dotted-dashed line), $d=1.7$ (dashed line), where the horizontal
line corresponding to $\Delta a^{}_e = 15 \times 10^{-12}$ is the
difference between the SM prediction and the experimental data.}
\end{figure}

\begin{figure}
\vspace{-12.0cm}
\epsfig{file=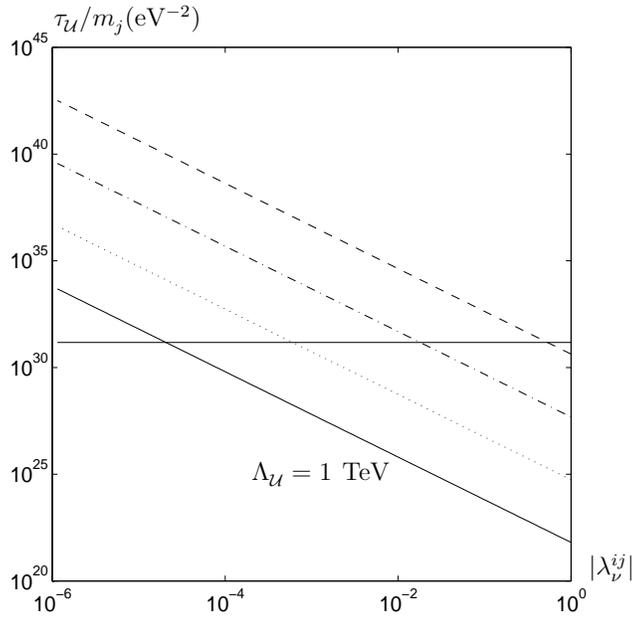,bbllx=0.5cm,bblly=14cm,bburx=12.5cm,bbury=34cm,%
width=10cm,height=16cm,angle=0,clip=0} \vspace{6.0cm}
\caption{Numerical illustration of $(\tau^{}_{\cal U}/m^{}_j,
|\lambda^{ij}_\nu|)$ for different scaling dimensions of the scalar
unparticle operator: $d = 1.1$ (solid line), $d=1.3$ (dotted line),
$d=1.5$ (dotted-dashed line), $d=1.7$ (dashed line), where the
horizontal line corresponds to the future cosmological bound $\tau/m
\geq 1.5 \times 10^{31}~{\rm eV^{-2}}$.}
\end{figure}

\end{document}